\documentclass[aps,prd,groupedaddress]{revtex4}
\usepackage{graphicx}
\usepackage{epsfig}
\usepackage{amsfonts}
\usepackage{amssymb}
\usepackage{epsf}
\usepackage{color}
\newcommand{\insertplot}[5]{\begin{figure}
 \hfill\hbox to 0.05in{\vbox to #5in{\vfill
 \inputplot{#1}{#4}{#5}}\hfill}
 \hfill\vspace{-.1in}
 \caption{#2}\label{#3}
 \end{figure}}
\newcommand{\inputplot}[3]{
 \special{ps: plotfile #1}
}

\newcounter{fig}

\newcommand{\ee}{\end{equation}}
\newcommand{\eea}{\end{eqnarray}}
\newcommand{\be}{\begin{equation}}
\newcommand{\bea}{\begin{eqnarray}}

\begin{document}

\title{Spherically symmetric charged black holes with wavy scalar hair}
\author{Yves Brihaye}
\email[]{yves.brihaye@umons.ac.be}
\affiliation{Service de Physique de l'Univers, Champs et Gravitation, Universit\'e de Mons, Mons, Belgium}
\author{Betti Hartmann}
\email[]{b.hartmann@ucl.ac.uk}
\affiliation{Department of Mathematics, University College London, Gower Street, London, WC1E 6BT, UK}

\date{\today}

\begin{abstract}
We study standard Einstein-Maxwell theory minimally coupled to a complex valued and self-interacting scalar field.
We demonstrate that new, previously unnoticed spherically symmetric, charged black hole solutions
with scalar hair exist in this model for sufficiently large gravitational coupling and sufficiently 
small electromagnetic coupling. The novel scalar hair 
has the form of a spatially oscillating ``wave packet'' and back-reacts on the space-time such that both the Ricci and the Kretschmann scalar, respectively, possess qualitatively similar oscillations. 
 
\end{abstract}
 
 \maketitle
 
\section{Introduction}

It is well known for some decades now that black holes can carry scalar field hair if the scalar field model is non-linear.
The first example of this type was given within the Skyrme model minimally coupled to graviy \cite{luckock_moss}. While the Skyrme model is often considered as an effective model in the context of nuclear physics, black holes that carry scalar hair can also be constructed in models inspired by high energy physics. The SU(2) Yang-Mills-Higgs model with the Higgs field in the adjoint representation (and hence being real) possesses hairy black holes \cite{Breitenlohner:1991aa, Breitenlohner:1994di} which can be thought of as black holes residing inside the core of magnetic monopoles. However, diverse no-hair conjectures
(see \cite{herdeiro2018asymptotically} for a recent review)  seem to prevent the existence of black holes with scalar hair charged under an abelian group. In recent years and motivated by an increased interest in the phenomenon of superradiance \cite{superradiance}, it was realized that black holes can carry complex valued scalar hair under certain conditions. The first (and best-known) example was put forward in the context of rotating, uncharged black holes
\cite{Hod:2012px, Herdeiro_2014}. The no-hair conjecture can be cirumvented in this case because of the  harmonic dependence on the time and azimuth coordinates and assuming the scalar field frequency to be fine-tuned to the horizon angular frequency of the black hole.
This so-called {\it synchronization condition} appears exactly at the threshold of superradiance and allows for so-called
{\it scalar Q-clouds} to exist on and close to the black hole horizon. 

Now, returning to models that contain U(1) gauge fields, a similar construction is possible -- even when the black hole is non-rotating
and spherically symmetric \cite{Herdeiro:2020xmb}. In this case, the frequency of the complex valued scalar field
needs to be fine-tuned to the electric potential on the horizon. It was then realized that when sufficiently strong back-reaction of the scalar cloud is taken into account that next to the expected extremal black holes with diverging
scalar field derivative on the horizon, a new type of solution exists that represents a black hole with an inflating exterior
\cite{Brihaye:2020vce}. On a shell of given thickness outside the horizon, the scalar field becomes constant and non-vanishing such that the scalar field potential energy corresponds to a positive cosmological constant. Moreover, the charge of these black holes gets screened. 

In this paper, we point out that additional solutions to the model studied in  \cite{Herdeiro:2020xmb,Brihaye:2020vce}
exist which have not been noticed so far and which have a novel type of scalar hair. 
These solutions exist at large gravitational coupling and small gauge field coupling and possess
``wavy'' scalar hair in the sense that the scalar field constitutes spatial oscillations outside the black hole horizon.
These oscillations lead to qualitatively similar oscillations in the curvature of the space-time. 
We will discuss these solutions and their properties in the following and point out that the phenomenon is independent of the choice of the self-interaction potential of the scalar field.

\section{The model}
We study a simple (3+1)-dimensional model in which Einstein gravity is minimally coupled to an electromagnetic field as well as to a complex valued, self-interacting scalar field. The Lagrangian density reads
\begin{equation}
\label{action}
{\cal{L}} = \frac{1}{16 \pi G} R   - \frac{1}{4} F^{\mu \nu} F_{\mu \nu} 
- D_{\mu} \Psi^{\dagger}  D_{\mu} \Psi - U(|\Psi|)  \ , 
\end{equation}
where $R$ is the Ricci scalar, $F_{\mu \nu}$ denotes the field strength tensor of the electromagnetic potential
$A_{\mu}$ and 
 $D_{\mu} \Psi = \left(\partial_{\mu} - i g A_{\mu}\right)\Psi$ denotes the covariant derivative of the scalar field $\Psi$. 
 The important point in our construction is that the scalar field is self-interacting. In the following, we will study two different potentials $U(|\Psi|)$. These read, respectively~:
 \begin{equation}
\label{poly}
 {\rm polynomial \ :} \ \ \ U_1(|\Psi|) = \mu^2 |\Psi|^2 - \lambda |\Psi|^4 + \nu |\Psi|^6 \ \ 
\end{equation}
and
 \begin{equation}
\label{expo}
 {\rm exponential \ :} \ \ \ U_2(|\Psi|) =  \mu^2 \eta^2 \left[1 - \exp\left(-\frac{\vert\Psi\vert^2}{\eta^2}\right)\right] \ .
\end{equation}
The parameter $\mu$ represents the rest mass of the scalar field.
The potential $U_1$ is motivated by studies of scalar fields in flat space-time and proves essential in the construction of scalar hair on charged black holes \cite{Herdeiro:2020xmb}. The potential $U_2$ appears in models describing
gauge-mediated supersymmety breaking \cite{susy1,susy2}. Of course, the polynomial potential $U_1$ is an approximation to the exponential potential $U_2$ for small scalar fields. Note, however, that the potential $U_1$ depends {\it a priori} on the independent parameters $\lambda$ and $\nu$ that can be freely chosen, while the potential $U_2$ is fully fixed by
choosing the mass of the scalar field as well as the energy scale $\eta$. 

In the following, we will discuss the simplest possible solution that can be constructed in this model: a static,
spherically symmetric, electrically charged black hole. The symmetries in the space-time as well as the U(1) gauge symmetry allow the following Ansatz for the metric, gauge field and scalar field, respectively: 
\begin{equation}
{\rm d}s^2 = -(\sigma(r))^2 N(r) {\rm d}t^2 + \frac{1}{N(r)} {\rm d}r^2 + r^2\left({\rm d}\theta^2 + \sin^2 \theta {\rm d}\varphi^2 \right) \ \ , \ \   A_{\mu} {\rm d} x^{\mu} = V(r) {\rm d} t  \ \ , \ \ 
\Psi=\psi(r) \exp(i\omega t) \ ,
\end{equation} 
where $\omega$ is a real-valued constant. 

As is usual in systems with a  U(1) gauge symmetry, the phase of the scalar field can be gauged away. Stating it differently, the resulting equations depend only the combination $\Omega:=\omega - g V(r=\infty)$. 
Now, it is well known that a charged black hole does not {\it per se} possess scalar hair, but that
a so-called {\it synchronization condition} has to be fulfilled in order for the black hole to carry a 
Q-cloud of scalar fields. This condition follows from the regularity of the 
matter fields on the horizon and reads $\omega-g V(r_h)=0$. In the following, we will choose the gauge $\omega=0$ 
and hence the boundary condition for the electric potential reads: $V(r_h)=0$ 

The equations of motion following for the variation of the action have been given in other publications (see e.g. \cite{Herdeiro:2020xmb, Brihaye:2020vce}) and we refer the reader to these papers. These equations depend on a number of dimensionless coupling constants
that result from appropriate rescalings. These are~:

\begin{itemize}
\item For the potential $U_1$ we use
rescalings $r\rightarrow r/\mu$, $v= \frac{\sqrt{\lambda}}{\mu} V$, $\psi = \frac{\sqrt{\lambda}}{\mu} \Psi$  such that the dimensionless
couplings read
\begin{equation}
      \alpha = \frac{4 \pi G \mu^2}{\lambda} \ \ , \ \ 
			\beta^2 = \frac{\nu \mu^2}{\lambda^2} \ \ , \ \ 
			e = \frac{g}{\sqrt {\lambda}}  \ . 
\end{equation}

\item For the potential $U_2$ we use the same rescaling for the radial coordinate as above, but now
$v= \frac{V}{\eta}$, $\psi = \frac{\Psi}{\eta}$  which results in the dimensionless coupling constants
\begin{equation}
      \alpha = \frac{8 \pi G \eta^4}{\mu^2} \ \ , \ \
			e = \frac{\eta}{\mu} g  \ . 
\end{equation}
\end{itemize}
We have chosen to use the same letters for the dimensionless couplings to demonstrate the qualitative similarily of our results for both potential.

From the equations of motion  we can infer the asymptotic behaviour of the functions, which reads
\begin{equation}
\label{eq:infty}
N(r \gg 1)=1-\frac{2M}{r} + \frac{\alpha Q_e^2}{r^2} + ..... \ \ , \ \ 
v(r\gg 1)=\Phi - \frac{Q_e}{r} + .... \ \ , \ \  \psi(r\gg 1)\sim \frac{\exp(-\mu_{\rm eff,\infty} r)}{r} + .... \ \ , \  \
\end{equation}
where the parameters $M$ and $Q_e$ denote the (dimensionless) mass and electric charge of the solution, respectively. 
Moreover,  $\mu_{{\rm eff},\infty}$ is the effective mass of the scalar field which is given by the bare mass (rescaled to unity with our choice of rescalings) and the asymptotic value of $v(r)$ denoted by $\Phi$, i.e. reads $\mu_{{\rm eff},\infty}=\sqrt{1 - \Omega^2}$, $\Omega \equiv e \Phi$. Hence, increasing (decreasing) $\Omega$ 
will decrease (increase) the effective mass of the scalar field. Note that the larger the effective mass of the scalar field, the stronger the scalar field will be localised. 

The scalar cloud (Q-cloud) surrounding the charged black hole can be thought of as made up of $Q_N$ scalar bosons, where $Q_N$ is the globally conserved Noether charge~:
\begin{equation}
\label{eq:noether}
Q_N=\int {\rm d} r \ \frac{2r^2 e v\psi^2}{N\sigma}  \ .
\end{equation}
Through the coupling to the electromagnetic field each of the scalar bosons carries a charge $e$. The total electric
charge $Q_e$ of the solution is then a sum of the electric charge in the cloud given by $eQ_N$ and the horizon electric charge $Q_H$ of the black hole given by $Q_H=v'(r_h) r_h^2/\sigma(r_h)$ \cite{Herdeiro:2020xmb} .
Next to the electric charge, we can also attribute a mass to the cloud. This is given by \cite{Herdeiro:2020xmb} ~:
\begin{equation}
M_{Q}=\frac{1}{4\pi} \int {\rm d}^3 x \ \sqrt{-g} \left(T^i_{\ i} - T^t_{\ t}\right) \ \ , \ \  i=1,2,3 \ .
\end{equation}

Finally, we the Hawking temperature of the black hole is given by 
\begin{equation}
 T_H=(4\pi)^{-1}\sigma(r_h) N'\vert_{r=r_h}   \ 
\end{equation}
and is related to the masses $M$ and $M_Q$ via the Smarr relation \cite{Herdeiro:2020xmb} ~:
\begin{equation}
2 T_H {\cal S} = M - \alpha M_Q \ \ , \ \  {\cal S}=\pi r_h^2 \ ,
\end{equation}
where ${\cal S}$ is the black hole entropy. 


\section{Black holes with novel scalar hair}
Charged black hole solutions with scalar hair of the type that we are suggesting here have been discussed previously in
 \cite{Herdeiro:2020xmb} and \cite{Brihaye:2020vce}, respectively. 
The system of equations has to be solved subject to regularity conditions on the horizon where $N(r_h)=0$ as well
as the requirement of asymptotic flatness and finite energy, respectively. While $v(r_h)=0$ is required for regularity, we have the choice to either fix the value of the electric field on the horizon $\propto v'(r_h)$ or the effective mass of
the scalar field by setting $\Phi$. In this paper, we will assume the former approach and show that when
fixing the gravitational back-reaction $\alpha$ we find that up to {\it three branches} of solutions exist when varying
$v'(r_h)$. The third branch, which has very peculiar new features as compared to the solutions on the other two branches,
had not been noticed before. We will also demonstrate in the following that this phenomenon appears for
both the polynomial as well as for the exponential potential, respectively.

\subsection{Potential $U_1$}
\begin{figure}[h!]
\begin{center}
{\includegraphics[width=5cm, angle=-90]{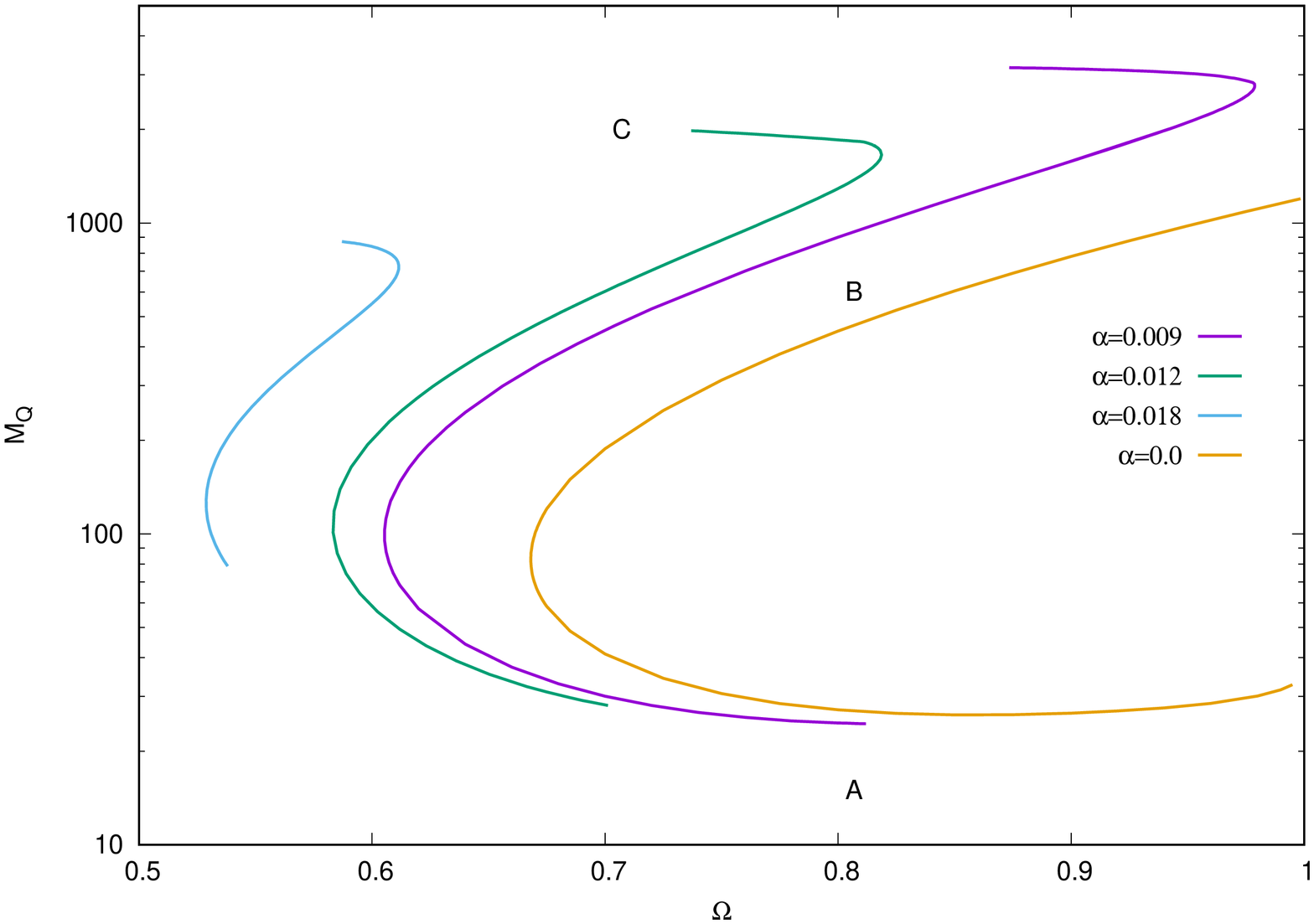}}
{\includegraphics[width=5cm,angle=-90]{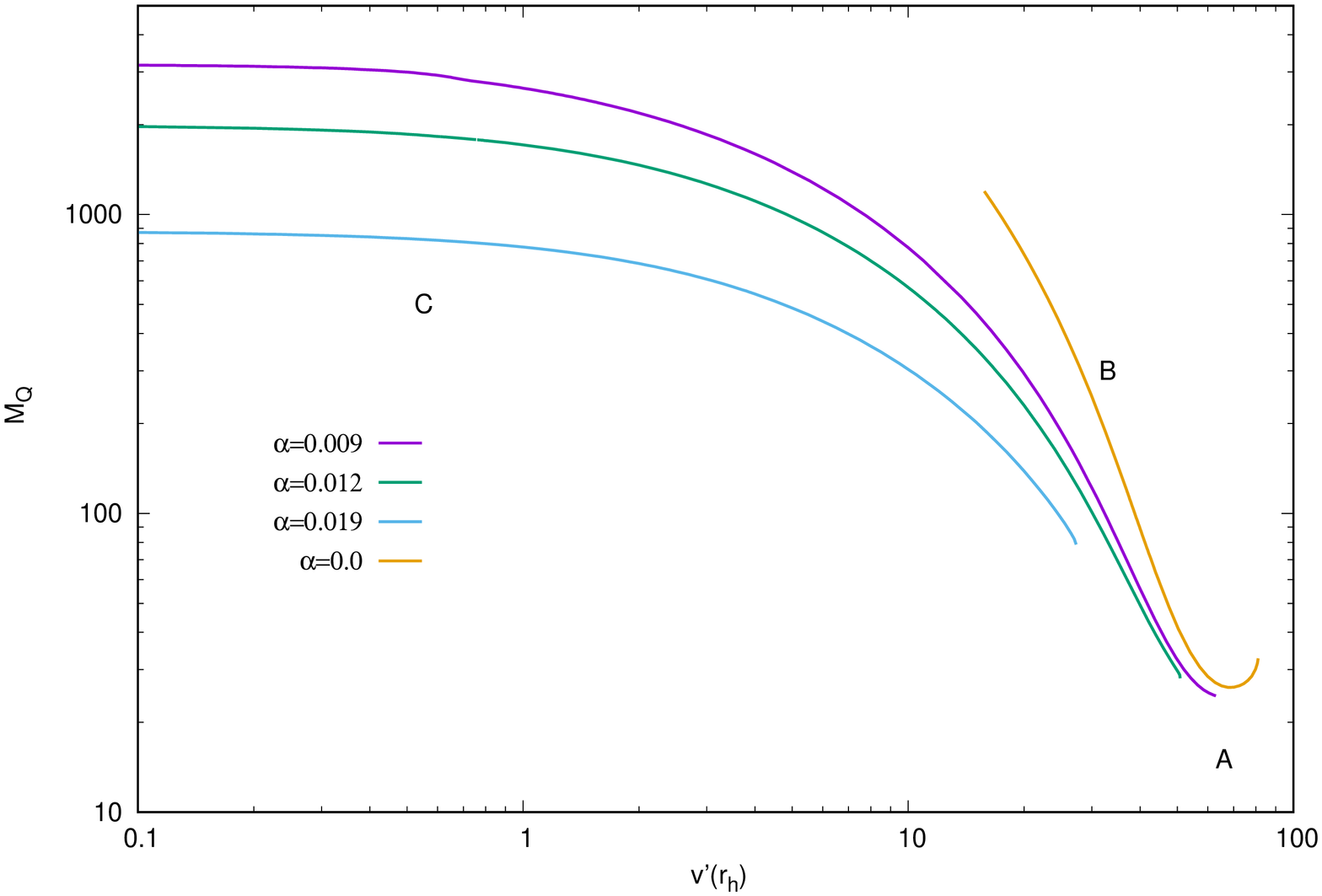}}
\caption{{\it Left}: 
The dependence of the mass $M_Q$ on the parameter $\Omega$ for different values of $\alpha$
with $e = 0.08$, $r_h = 0.15$, $\beta = 9/32$. 
{\it Right}: The  dependence of  $M_Q$ on $v'(r_h)$ for the same set of solutions.
\label{fig:fig_MQ}
}
\end{center}
\end{figure}

In order to understand the pattern of solutions, we have first decoupled the matter fields from the space-time geometry, i.e. we have studied the case $\alpha=0$ and  fixed  $\beta$, $r_h$ and  $e > 0$.   When varying the electric field on the horizon $\propto v_h'$, we find the well known result that 
two branches of solutions of scalar Q-clouds exist in $\Omega \in [\Omega_{\rm min}, 1]$, where the upper value of the interval
results from the choice of potential parameters, while the value of $\Omega_{\rm min}$ depends on the choice of 
$\beta$, $r_h$ and  $e > 0$. This is shown in Fig. \ref{fig:fig_MQ} for  $\beta=9/32$, $r_h=0.15$ and  $e=0.08$, where our numerical results indicate that $\Omega_{\rm min} \approx 0.668$. The two solutions with equal values of $\beta$, $r_h$, $e$ and $\Omega$ differ in the mass $M_Q$ and Noether charge $Q_N$. 
The branch of solutions with  lower (resp. higher) value of mass $M_Q$ is labelled ``A'' (resp. ``B'') and the two branches join at $\Omega = \Omega_{\rm min}$. On branch A, the mass of the scalar cloud is increased by adding scalar bosons.
The Noether charge $Q_N$ (not given here) shows a qualitatively very similar behaviour to $M_Q$. Now, scalar bosons can be added
up to a maximal effective mass of the scalar bosons, corresponding to a minimal $\Omega$. Then, adding more scalar bosons requires a decrease of the effective mass (branch B) and the maximum possible mass of the cloud is reached 
at $\Omega=1$ on branch B. It is also interesting to note that the increase in the number of scalar bosons
making up the cloud leads to a decrease in value of the electric field on the horizon, see Fig.\ref{fig:fig_MQ} (right).
We find that $v'(r_h)\approx 81.1$ for a solution on branch A close to $\Omega=1$, while it decreases to
$v'(r_h) \approx 15.7$ for the cloud with the larges possible $M_Q$ and $Q_N$. 

\begin{figure}[h!]
\begin{center}
{\includegraphics[width=5cm,angle=-90]{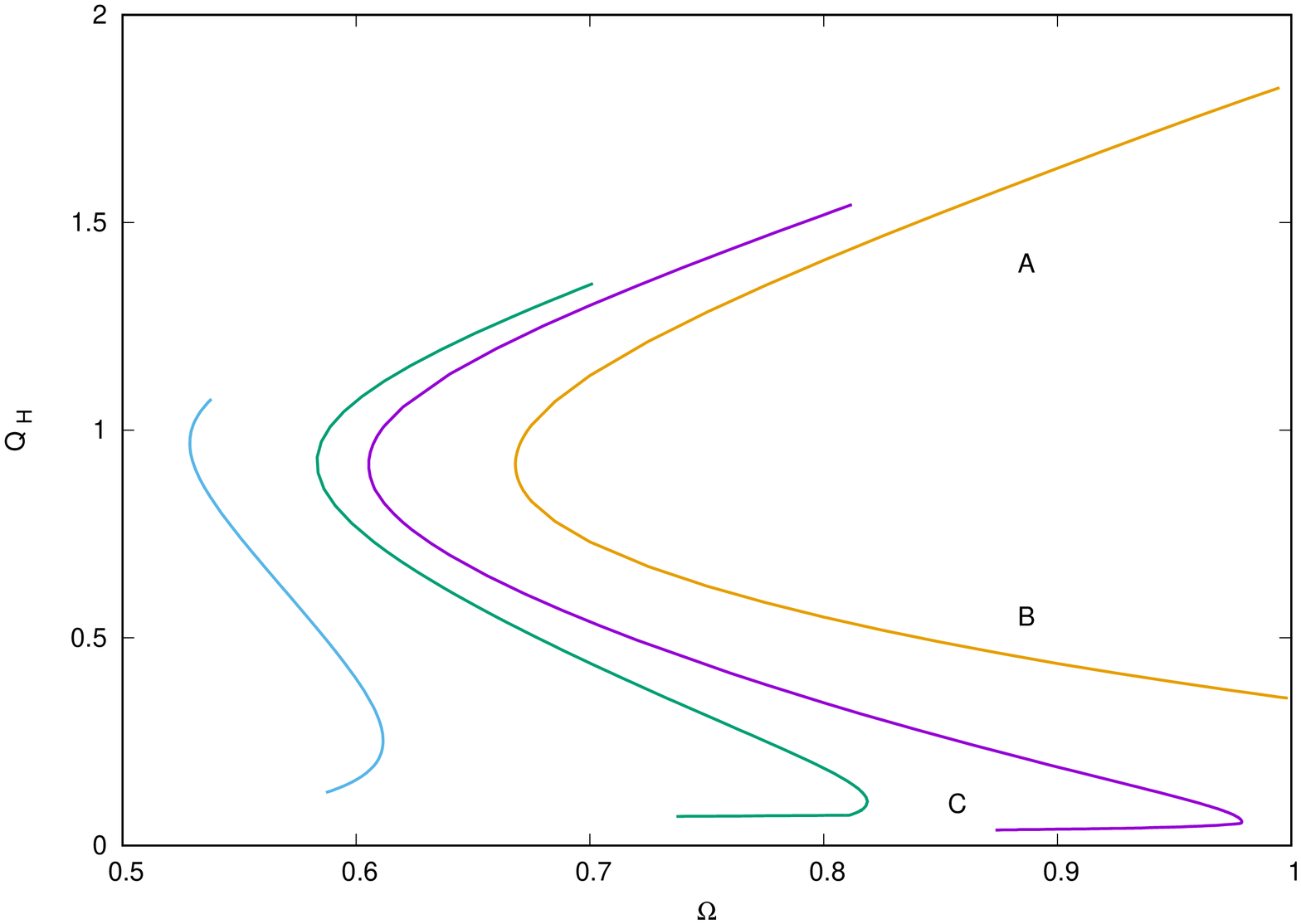}}
{\includegraphics[width=5cm, angle=-90]{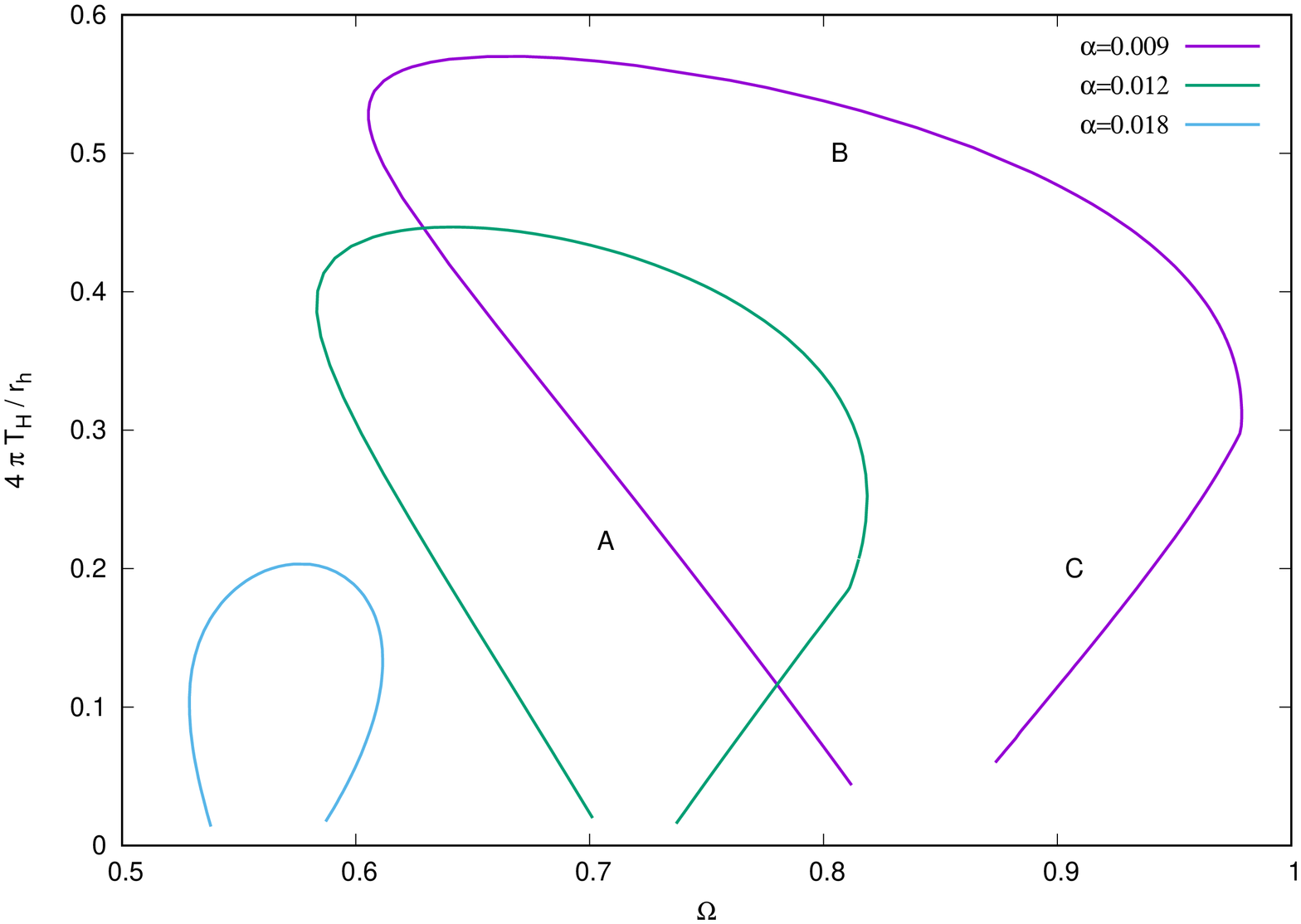}}
\caption{{\it Left}: The  dependence of the horizon electric charge $Q_H$ on $\Omega$ for the solutions
shown in Fig.\ref{fig:fig_MQ}. 
{\it Right}:  The dependence of the Hawking temperature $T_H$  on $\Omega$ for 
the solutions shown in Fig.\ref{fig:fig_MQ} for $\alpha > 0$.
\label{fig:QH_and_T}}
\end{center}
\end{figure}

Now, letting the scalar cloud back-react on the space-time, we find that the branches A and B
exist on modified intervals of $\Omega$, see Fig.\ref{fig:fig_MQ} (left). This is well understood and has been studied in
\cite{Herdeiro:2020xmb} and \cite{Brihaye:2020vce}, respectively. The new feature in the back-reacting case that has not
been noticed so far is that a third branch of solutions in $\Omega$ exists (labelled ``C'' in the following). 
On this branch, the mass $M_Q$ as well as the Noether charge $Q_N$ increase further, but now decreasing $\Omega$.
At the same time, the value of $v'(r_h)$ decreases to very small values, eventually reaching value zero at the end of branch C (see Fig.\ref{fig:fig_MQ} (right)). To state it differently: at the end of branch C the horizon electric charge $Q_H$ of the black hole becomes very small. This is shown in Fig. \ref{fig:QH_and_T} (left).  This means that all the electric charge is now in the cloud and the black hole is essentially ``discharged''.

Interestingly, the Hawking temperature $T_H$ of the black holes on this branch C shows a qualitatively different dependence
on $\Omega$ than branch A and B. This is shown in  Fig. \ref{fig:QH_and_T} (right), where
we give $T_H$ in function of $\Omega$. Obviously, on branch A and B, the black hole temperature increase when decreasing $\Omega$, while on branch C, the temperature of the black hole decreases with decreasing $\Omega$.
Now, remembering that along the branches from A to C, we increase the mass $M_Q$ and Noether charge $Q_N$, 
this means that while on branch A the addition of scalar bosons to the cloud (and hence the increase of mass
$M_Q$) leads to an increase of the temperature, the (further) addition of scalar bosons to the cloud leads to a decrease in
temperature on branch B and C. On branch B this happens for decreasing effective mass of the scalar field, while on branch C the temperature decreases with increasing $M_Q$ and increasing effective scalar boson mass.

\begin{figure}[h!] 
\begin{center}
{\includegraphics[width=5cm, angle=-90]{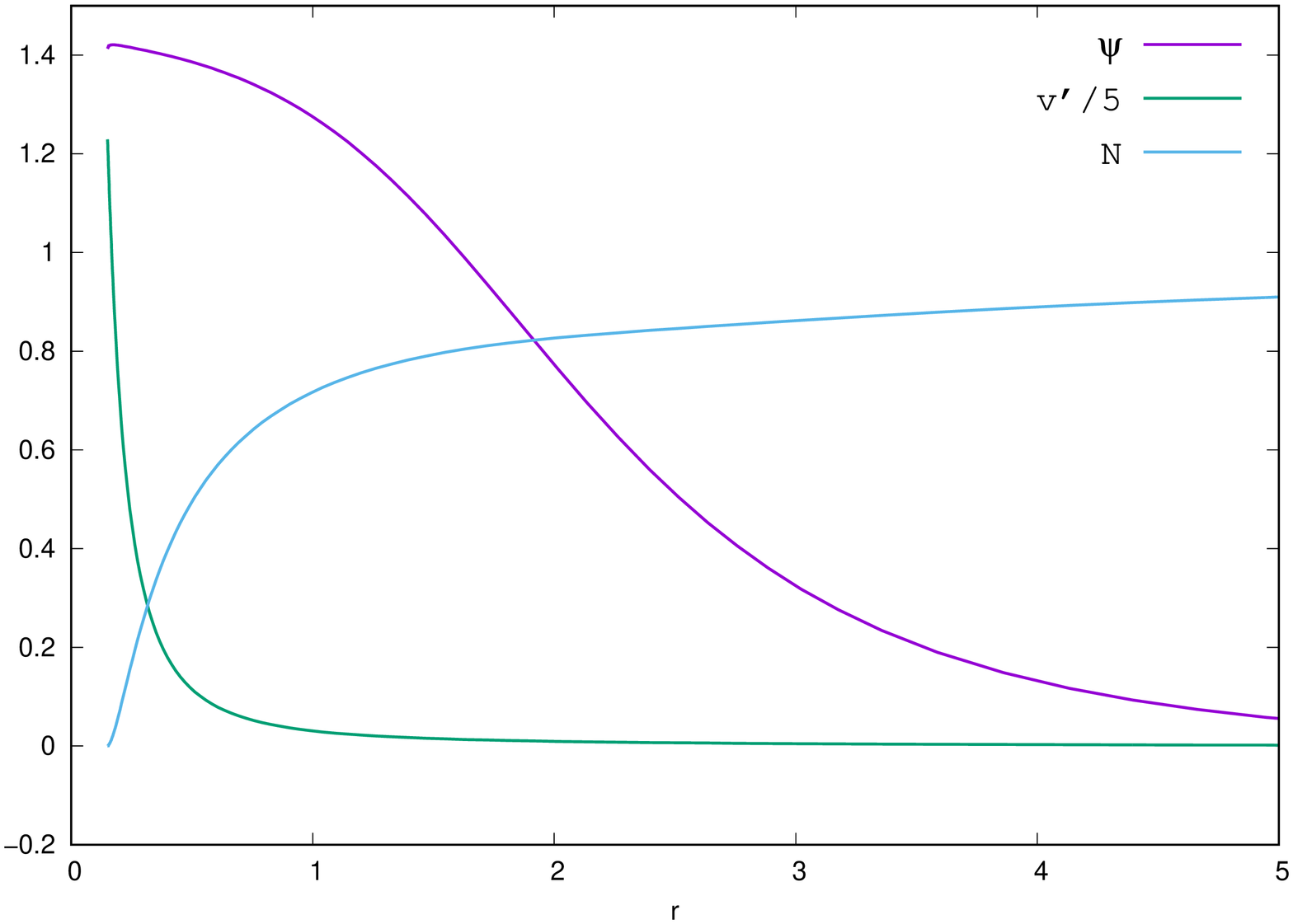}}
{\includegraphics[width=5cm,angle=-90]{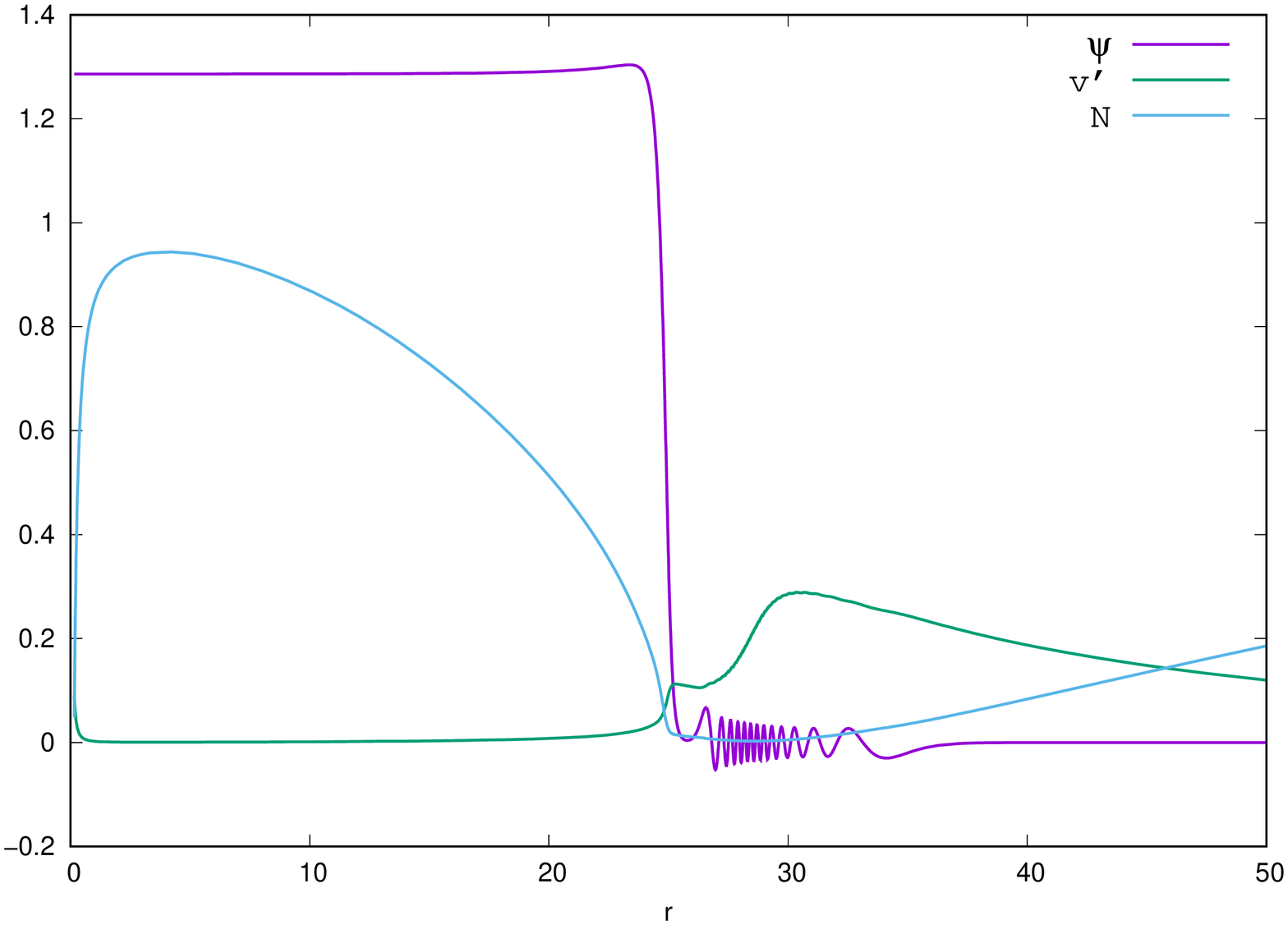}}
\caption{{\it Left}: The profiles of $\psi$ , $v'$ and $N$ for a solution on branch A with $\alpha = 0.09$, $e=0.08$, $r_h=0.15$ 
and $v'(r_h)\approx 60.0$.
{\it Right}: The profiles of $\psi$ , $v'$ and $N$ for a solution on branch C with the same values of $\alpha$, $e$ and $r_h$, but  $v'(r_h)= 0.1$.
\label{fig:profiles}
}
\end{center}
\end{figure}

Considering these physical parameters, the question is what distinguishes the solutions
on branch C from solutions on branch A and B. This is indicated in Fig. \ref{fig:profiles}, where we show the profiles
of the scalar field function $\psi(r)$, the electric field $\sim v'(r)$ as well as the metric function $N(r)$ for
a solution on branch A (left) and a solution on branch C (right). For both solutions, we have chosen $\alpha = 0.09$, $e=0.08$ and $r_h=0.15$, but the solution on branch A has $v'(r_h)\approx 60$, while the solution on branch C has $v'(r_h)\approx 0.1$. 

\begin{figure}[h!]
\begin{center}
{\includegraphics[width=5cm, angle=-90]{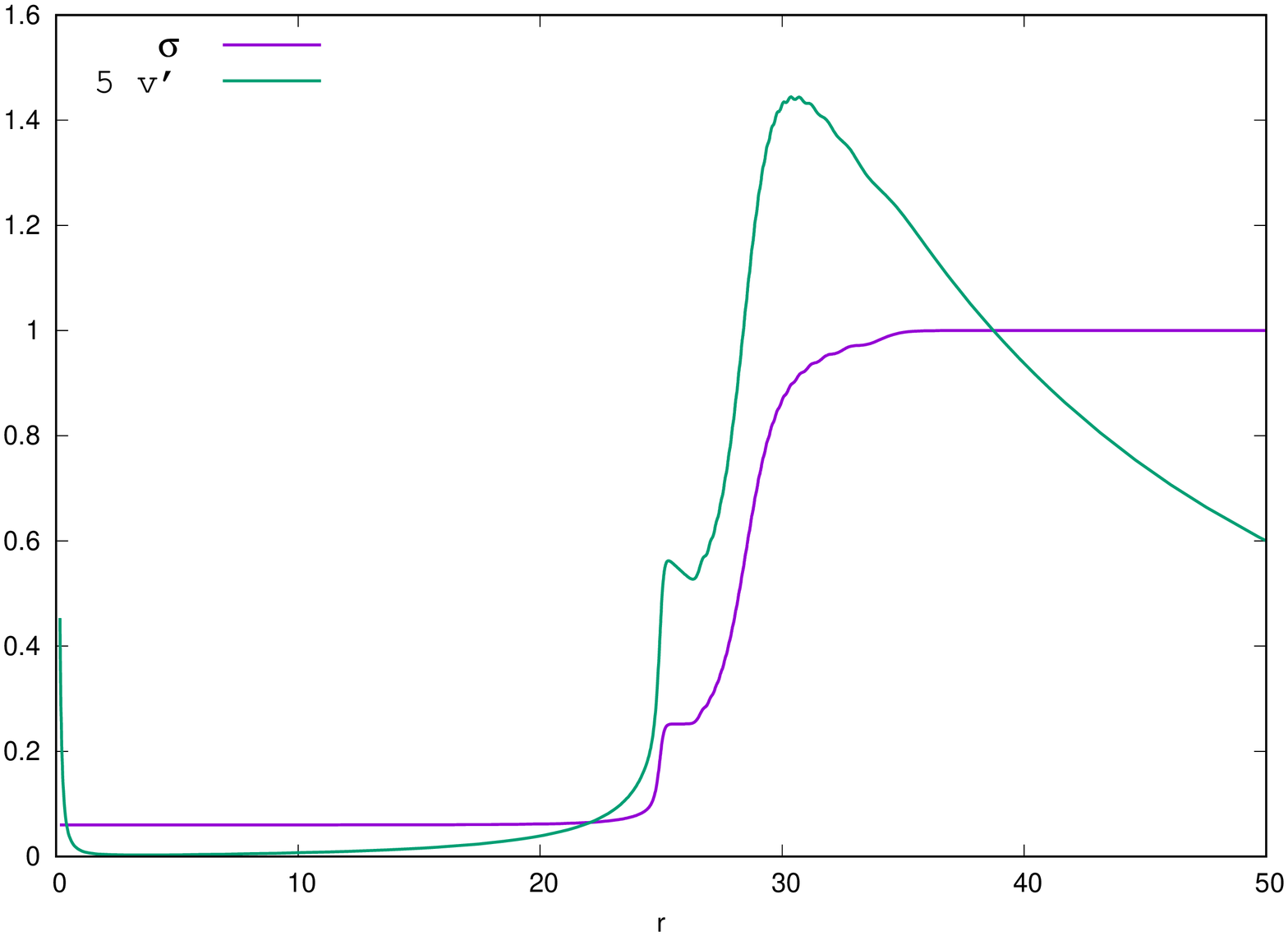}}
{\includegraphics[width=5cm,angle=-90]{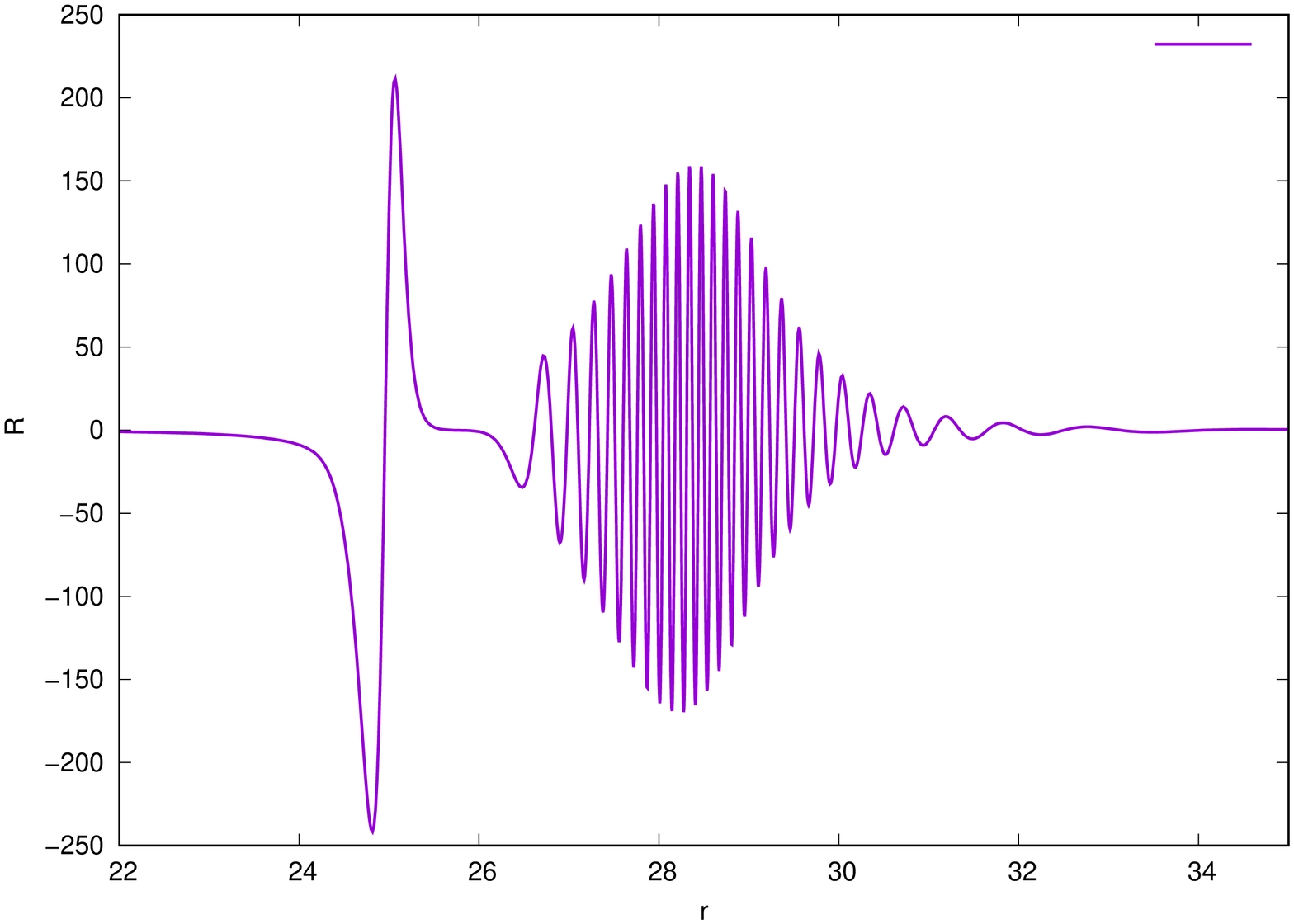}}
\caption{{\it Left}: Functions $v'(r)$ and $\sigma(r)$ for a solution on branch C  with $\alpha = 0.09$, $e=0.08$, $r_h=0.15$. {\it Right}: The corresponding Ricci scalar $R$ for this solution. 
\label{fig:profiles_and_R_branchC}
}
\end{center}
\end{figure}

Obviously, for the solution on branch A (similarly for a solution on branch B)
the scalar function $\psi(r)$ is a monotonic function of the radial variable~: $\psi(r)$ decreases
from its maximal value $\psi(r_h)$ to zero at infinity, where the fall-off is determined by
the effective mass, i.e. by $\Omega$ (see discussion above). Similarly, the function $v'(r)$ determining the electric field of the solution decreases monotonically and $N(r)$ increases monotonically from its value zero on the horizon to 
unity asymptotically.  This is very different for a solution on branch C. Here, the scalar field is nearly constant (and non-vanishing), while $v'(r)$ is very close to zero on an interval $r \in [r_h,r_c]$. The value $r_c > r_h$ denotes the approximate value of the radial coordinate at which the metric function $N(r)$ attains very small (but non-vanishing) values. Interestingly, instead of forming an extremally charged black hole (which is a solution to the model
for $\psi\equiv 0$), the non-linear interaction between the curvature of space-time, the electric field and the scalar field
now leads to spatial oscillations in the scalar field around zero on a finite interval of the radial coordinate. These spatial oscillations
are smaller in the other fields, but are present, as can been seen for an amplified $v'(r_h)$ and the metric function
$\sigma(r)$, respectively, in Fig. \ref{fig:profiles_and_R_branchC} (left). In particular, the solutions have significant amplitude
in the scalar curvature as given by the Ricci scalar $R$, see Fig. \ref{fig:profiles_and_R_branchC} (right). We do not give the Kretschmann scalar $K=R_{\mu\nu\rho\sigma} R^{\mu\nu\rho\sigma}$ here, but would like to state that the oscillations are also
present for $K$. This makes us believe that these are genuine oscillations that are not an artefact of the choice of coordinates. 

After a finite number of oscillations which decrease in amplitude for increasing $r$, the scalar field becomes identically
zero for $r > r_a$. The solution for $r\in (r_a: \infty)$ clearly correspond to a Reissner-Nordstr\"om solution~: $\sigma\equiv 1$ and the Ricci scalar $R\equiv 0$, while the electric field $\sim Q/r^2$. We hence find a novel black hole solution which possesses 
\begin{itemize}
\item an inflating exterior for $r\in [r_h:r_c]$ -- a phenomenon that
was already noticed and discussed in \cite{Brihaye:2020vce}, 
\item ``wavy'' scalar hair on a intermediate interval  $r\in (r_c:r_a]$ with metric function $N(r)$ close to zero, and
\item vanishing scalar field with the electric and metric fields showing a Reissner-Nordstr\"om behaviour for  
$r\in (r_a:\infty)$.
\end{itemize}

Let us insist here that the oscillations in the fields are {\bf not} a numerical artefact. We did perform several checks in this direction, e.g. we verified the independence of the solution on the interval of integration and/or the given tolerance.
Note that since our numerical integrator \cite{colsys} uses an adaptive grid scheme, changing the interval of integration
implies also the change of the discrete points at which the equations are evaluated. Moreover, we find full branches of
solutions that show a continuous dependence on the parameters. Also, because the new branch of solutions
is connected to the already known branches, we don't think that these solutions are
radially excited solutions similar to those observed in many non-linear models that possess soliton and/or black hole solutions.
These latter type of solutions typically would appear as new, disconnected branches with physical parameters
very different in value.

\begin{figure}[h!]
\begin{center}
{\includegraphics[width=5cm, angle=-90]{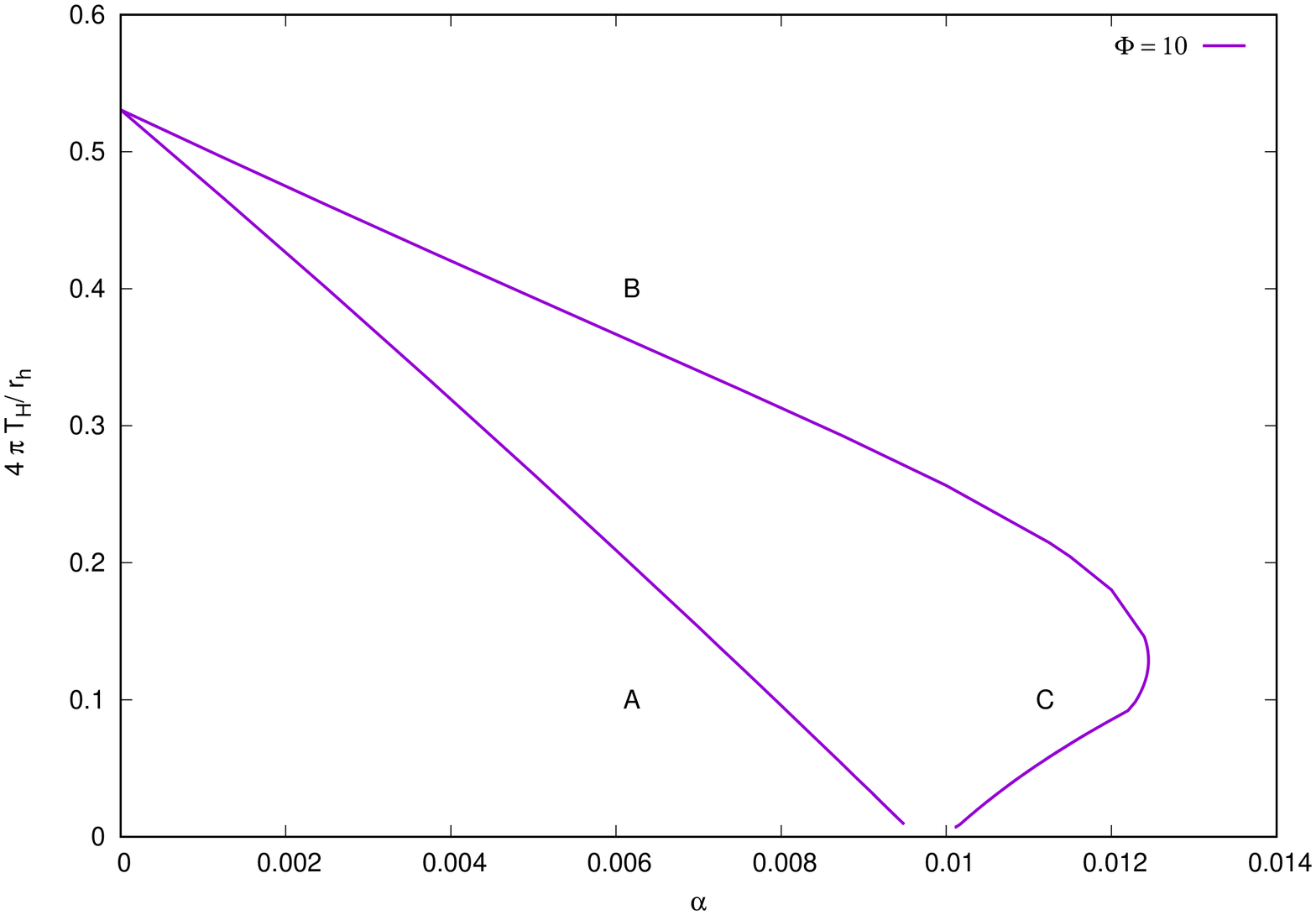}}
{\includegraphics[width=5cm,angle=-90]{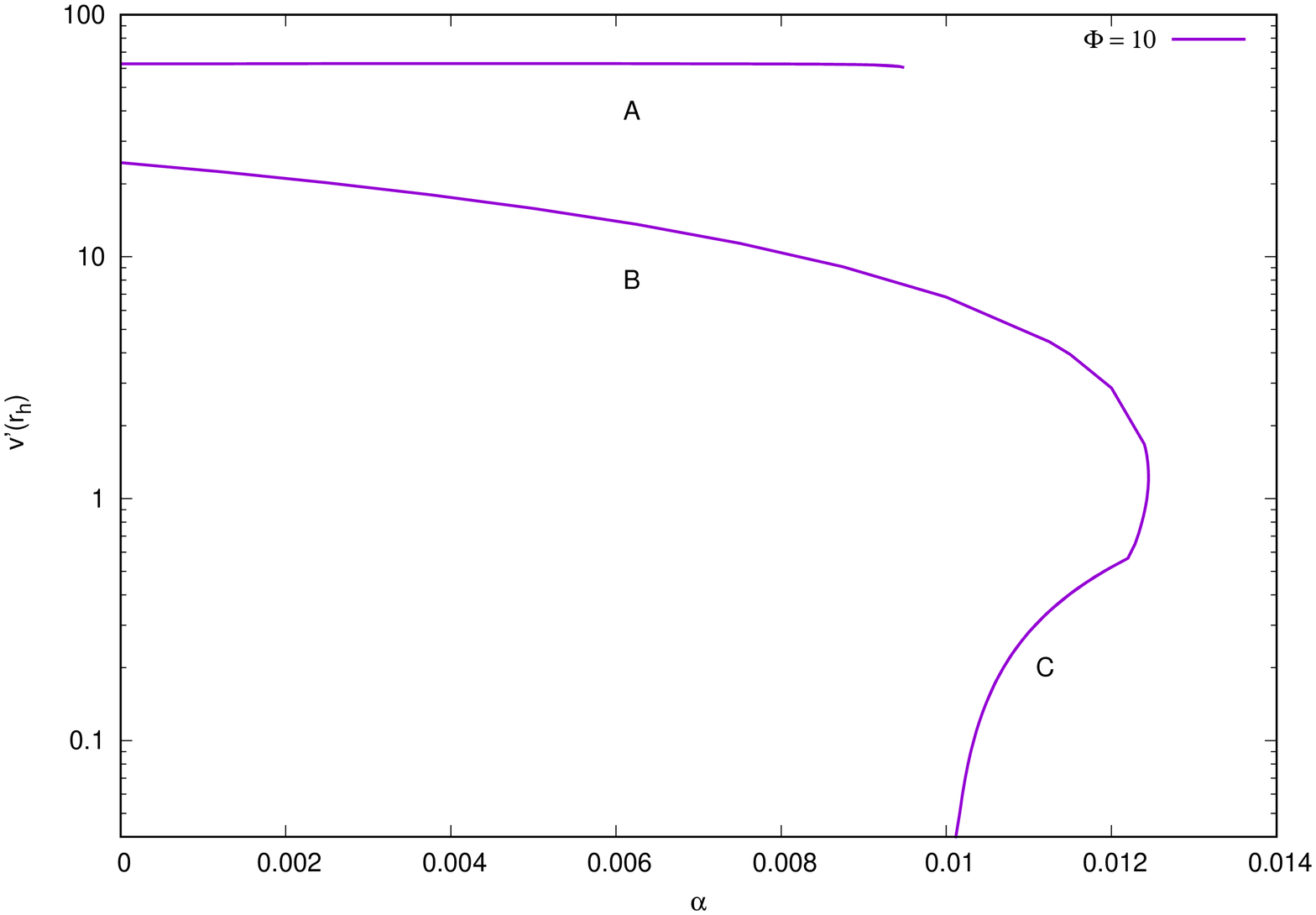}}
\caption{{\it Left}: We show the temperature of the black holes, $T_H$, in dependence on $\alpha$ for  $r_h=0.15$, $e=0.08$ and $\Omega = 0.8$ (corresponding to $\Phi=10$).
{\it Right}: The  $\alpha$-dependence of the electric field on the horizon $\sim v'(r_h)$ for the same family of solutions.
\label{fig:phi10}
}
\end{center}
\end{figure}

In order to understand these new solutions further, we have also studied the case of varying back-reaction and found up to three branches -- consistent with the above analysis. 
An example is shown in Fig.\ref{fig:phi10} for fixed $r_h=0.15$, $e=0.08$ and $\Omega=0.8$ (corresponding to $\Phi=10$). Here, we give the
temperature $T_H$ (left) as well as the electric field on the 
horizon $\sim v'(r_h)$ (right) in dependence on $\alpha$.  As is obvious from this figure, branch C is an extension of
branch B, but -- as mentioned above -- has different properties. E.g. while on branch B, the temperature decreases with $\alpha$, it increases with $\alpha$ on branch C. Branch A and B are connected to each other at $\alpha=0$, while
branch A and C seem to start, respectively end more or less at $T_H=0$. Hence, we find a (nearly) closed curve in the $\alpha$-$T_H$-plane.  Fig.\ref{fig:phi10} (right) also demonstrates that there exists a gap in $v'(r_h)$ for which no 
black hole solutions with scalar field exist at all. Increasing $\alpha$, this gap increases slightly. 

\subsection{Potential $U_2$}
\begin{figure}[h!]
\begin{center}
{\includegraphics[width=5cm, angle=-90]{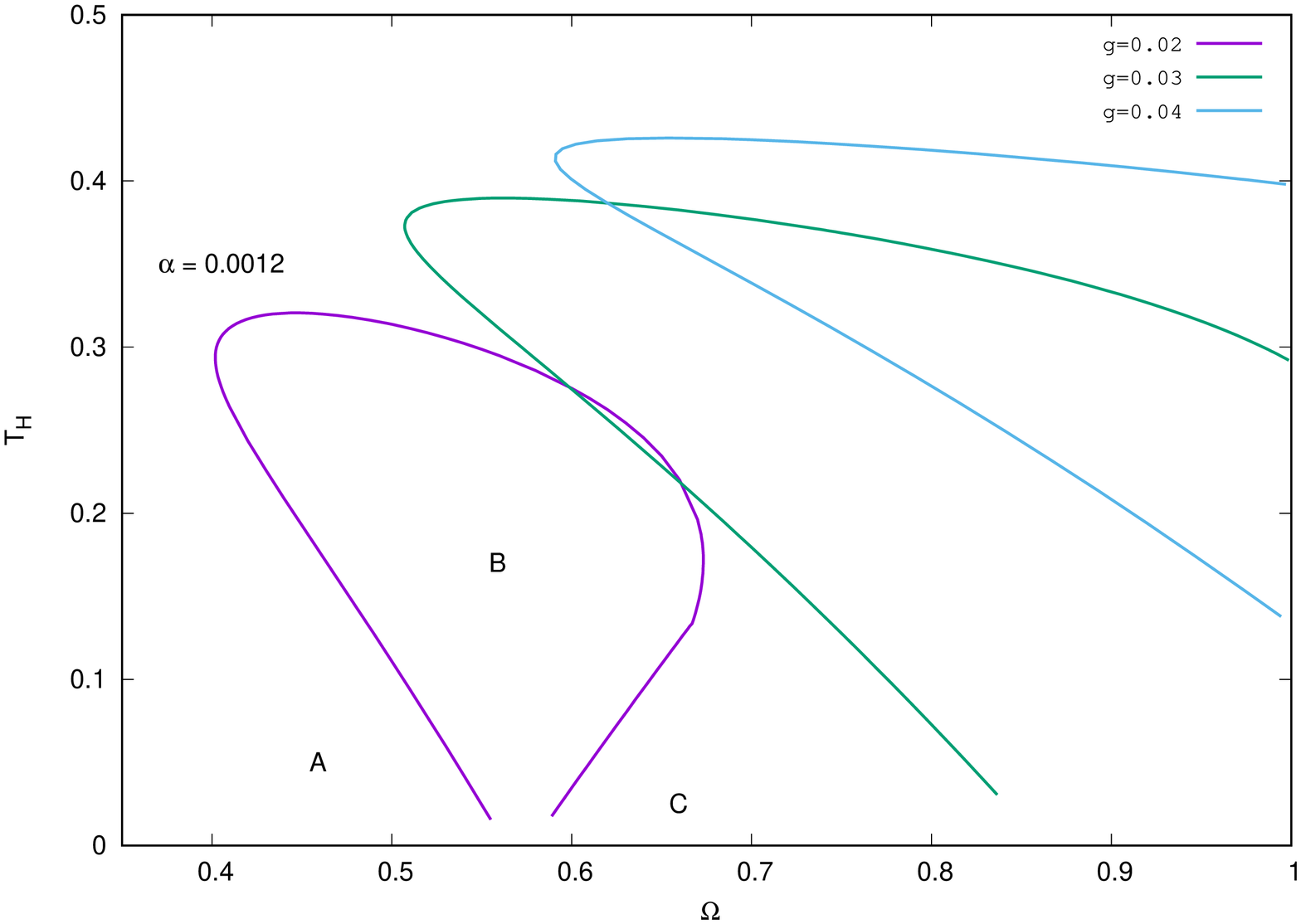}}
{\includegraphics[width=5cm,angle=-90]{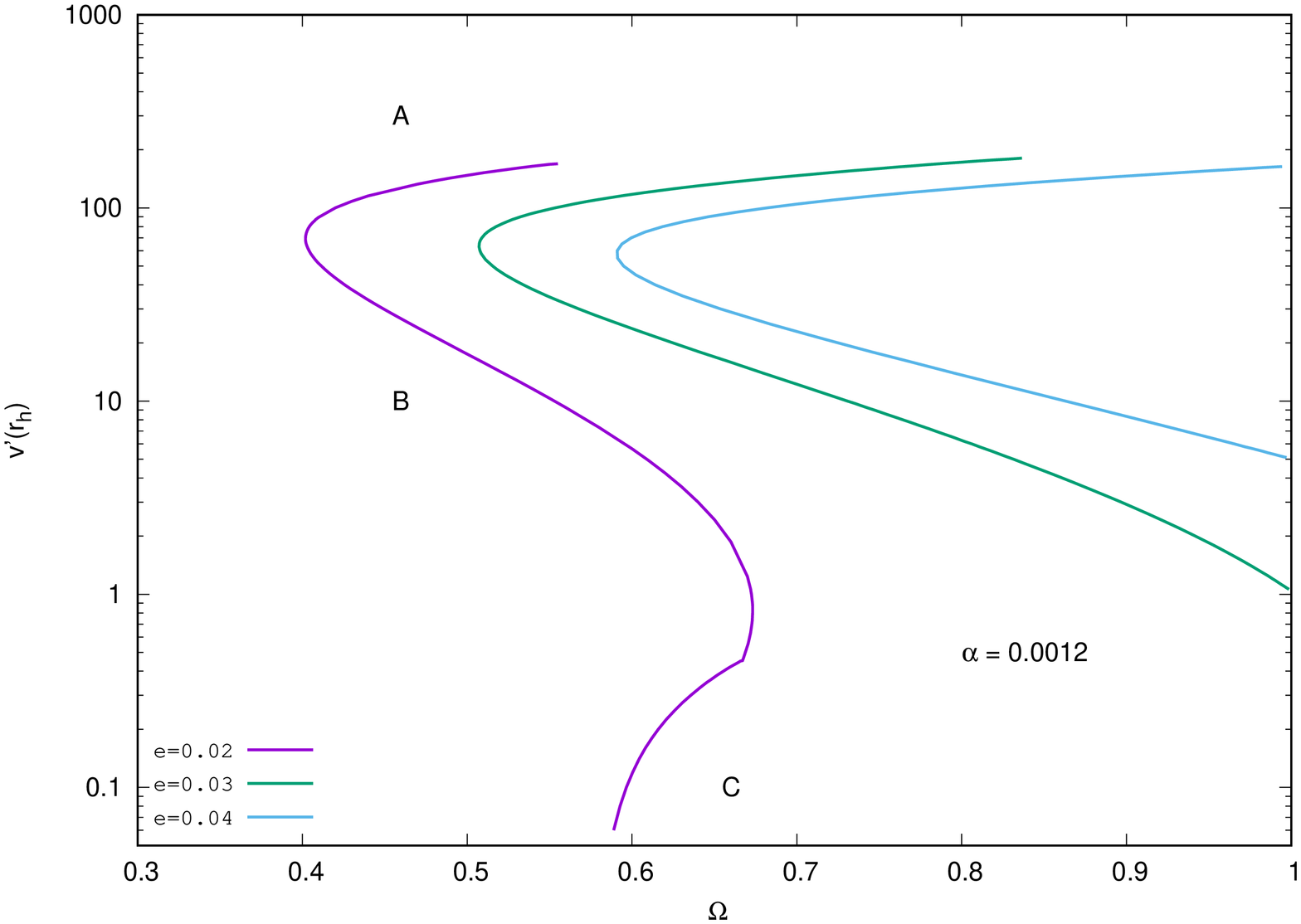}}
\caption{{\it Left}: We show the dependence of the temperature $T_H$ of the black holes on $\Omega$ for  
$\alpha=0.0012$, $r_h=0.15$ and different values of $e$. 
{\it Right}: Same as left for the electric field on the horizon $\sim v'(r_h)$. 
\label{fig:susy1}
}
\end{center}
\end{figure}

Studying the potential $U_2$, which is {\it a priori} better motivated physically as it appears in certain gauge-mediated supersymmetry breaking models, we find that the qualitative phenomenon of a third branch of solutions 
consisting of black hole solutions with wavy scalar hair persists. In the following (and not to repeat ourselves), we will put the emphasis on additional features of the solutions, in particular, we want to emphasize the role of the gauge coupling
$e$ here.  For that we have fixed $\alpha$ and varied $e$. Our results are given in  Fig. \ref{fig:susy1} for
$\alpha=0.0012$ and $r_h=0.15$. This figure demonstrates that the phenomenon described above
appears only for values of $e$ sufficiently small. For large $e$, branch B extends all the way back to $\Omega=1$, where
it terminates. In this latter case, the black hole has temperature significantly larger than zero and 
$v'(r_h) > 0$. Only when $e$ is small enough are we able to find branch C. For our choice
of parameters, only $e=0.02$ (in comparison to $e=0.03$ and $e=0.04$) allows the existence of this new branch.

\begin{figure}[h!]
\begin{center}
{\includegraphics[width=5cm, angle=-90]{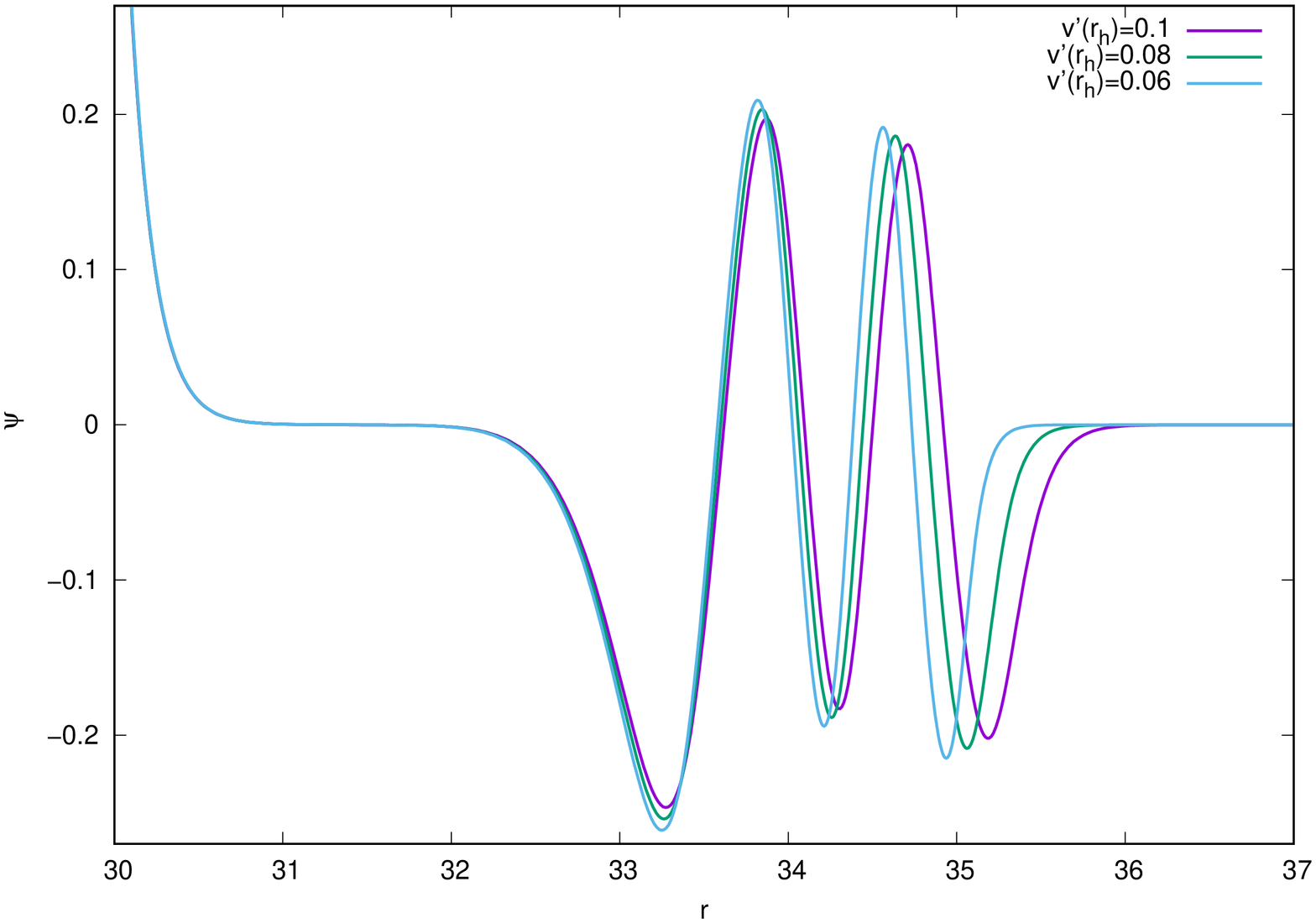}}
{\includegraphics[width=5cm,angle=-90]{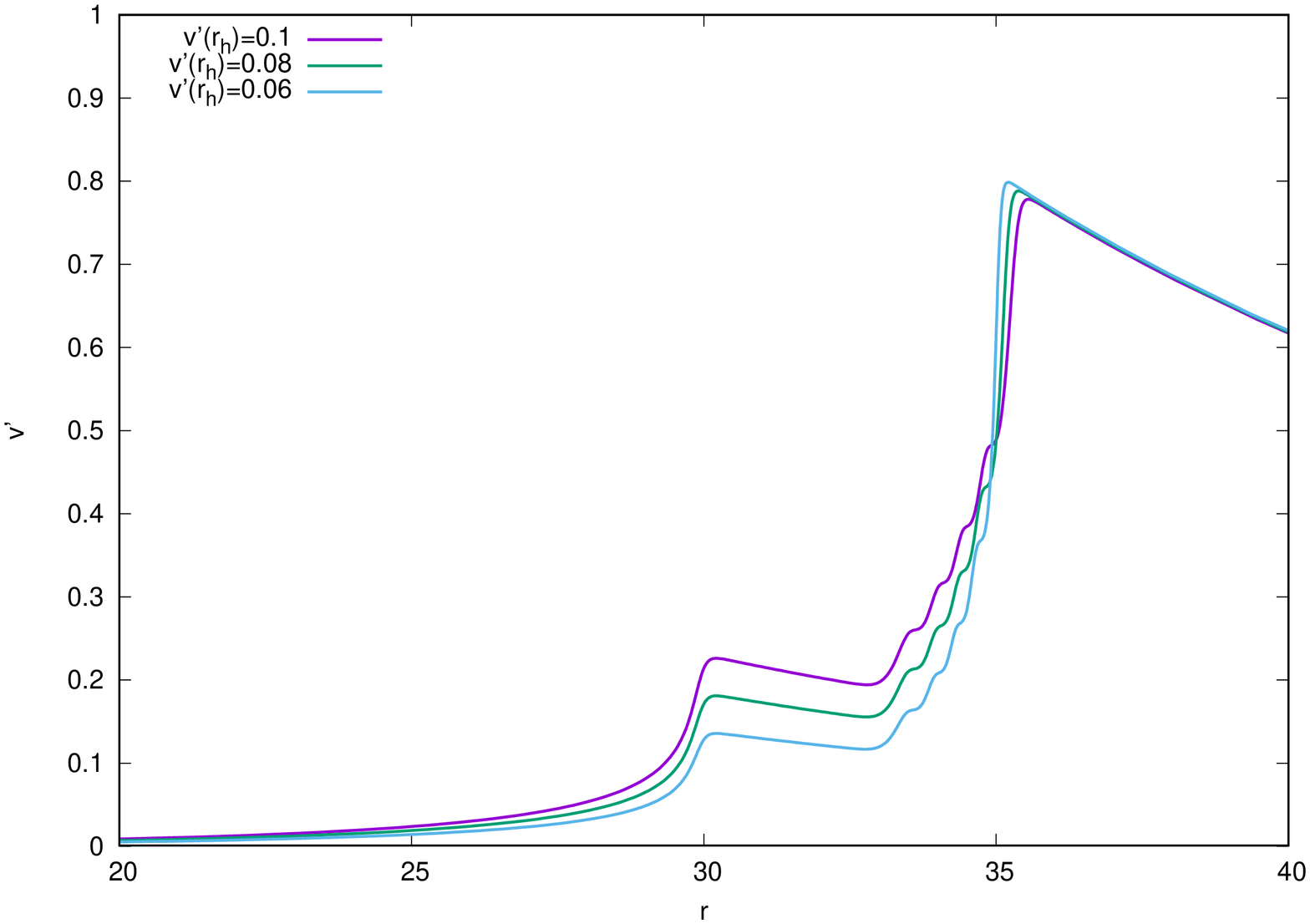}}\\
{\includegraphics[width=5cm, angle=-90]{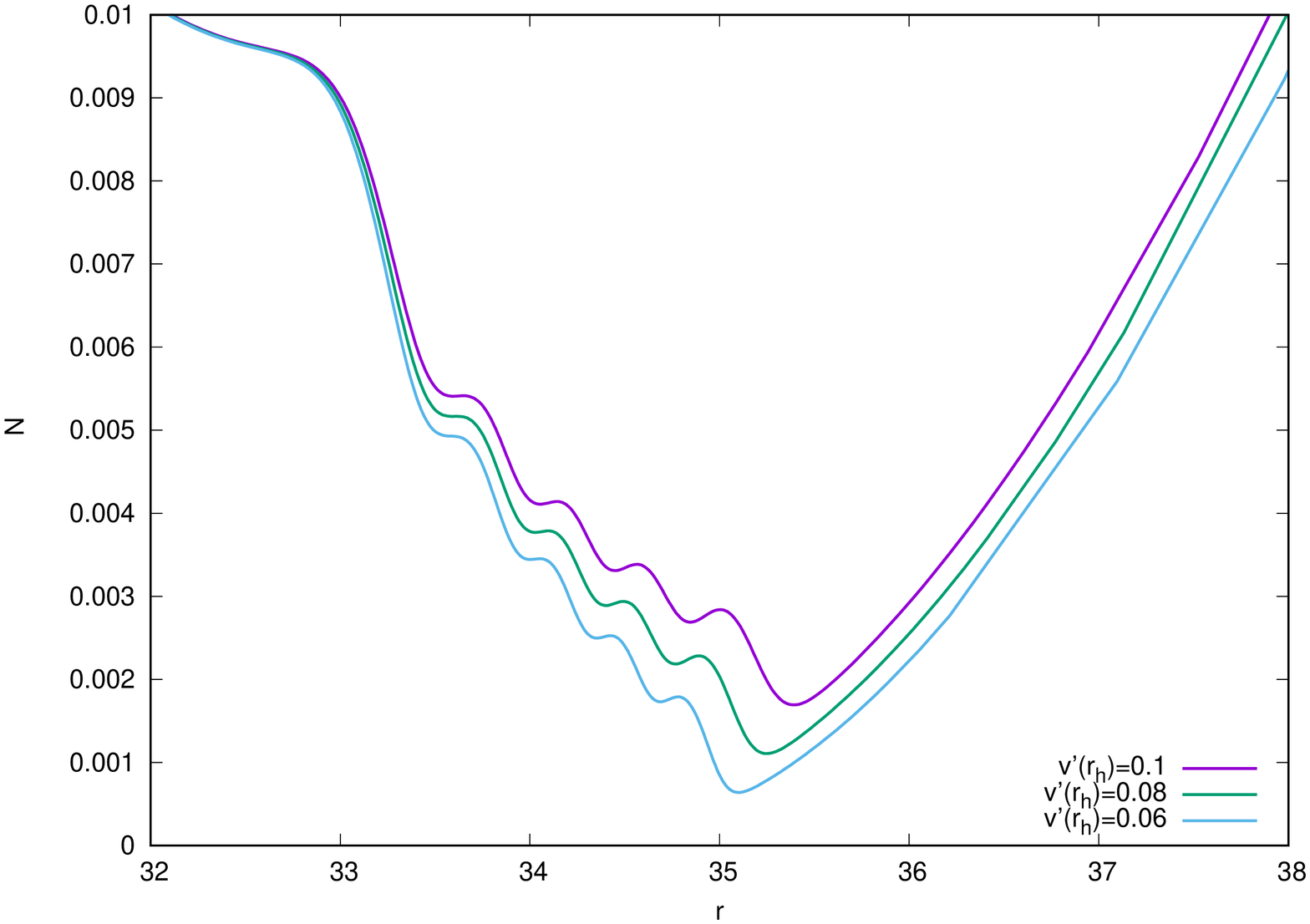}}
{\includegraphics[width=5cm,angle=-90]{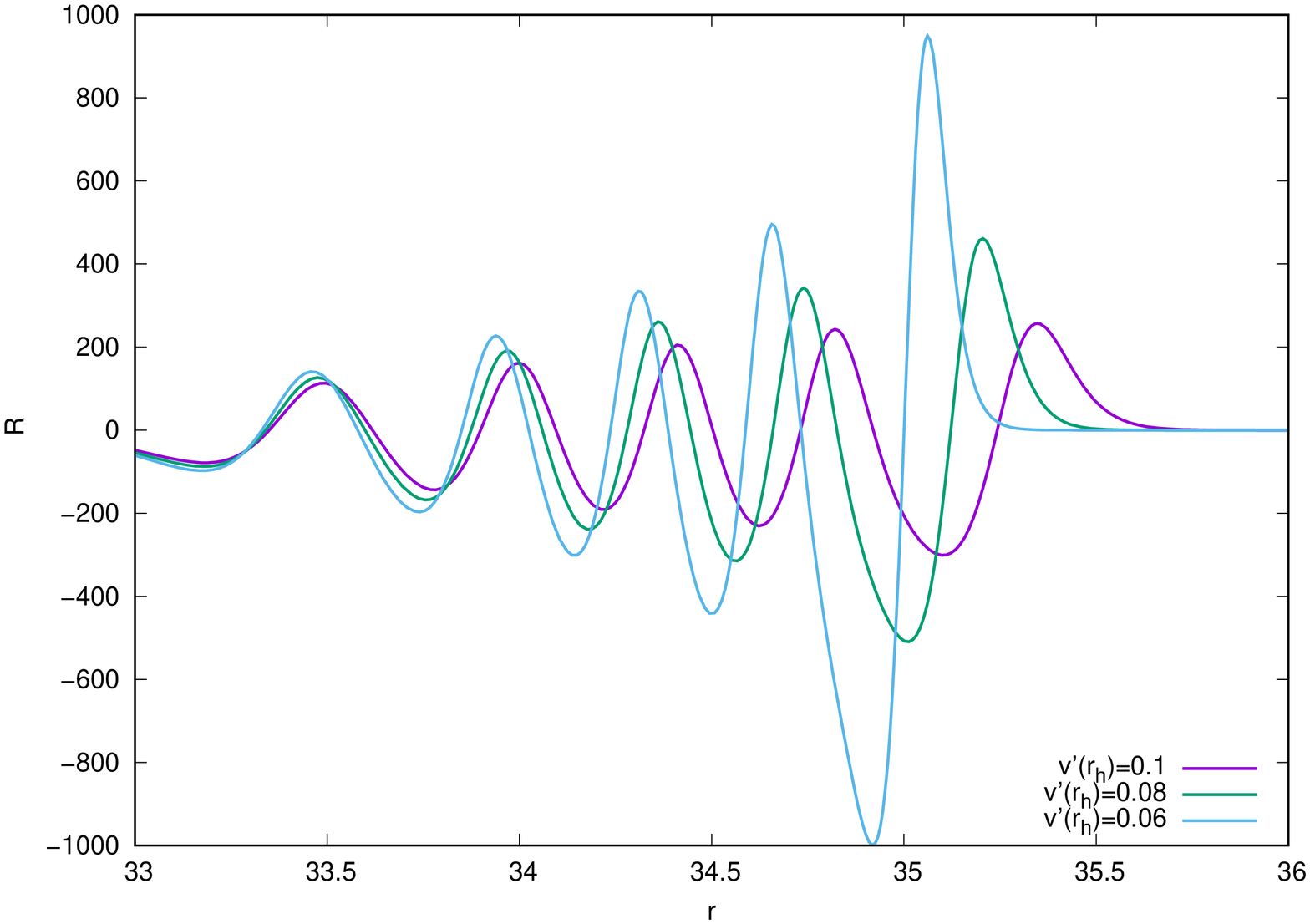}}
\caption{We show the evolution of solutions on branch C for $\alpha = 0.0012$, $e=0.02$ 
when varying $v'(r_h)$: profile of $\psi(r)$ (top left), 
profile of $v'(r)$ (top right), profile of $N(r)$ (bottom left) and profile of the Ricci scalar $R$ (bottom right). 
Note that we are only showing the interval of $r$ where oscillations of the respective profiles appear.
\label{fig:profile_susy_vp}
}
\end{center}
\end{figure}

Knowing the qualitative behaviour of the solutions on branch C, we have then checked how the solutions
evolve along this branch C.  So, in Fig. \ref{fig:profile_susy_vp}, we show the profiles of the functions 
as well as the Ricci scalar $R$ in the interval of $r$ where oscillations appear for $\alpha = 0.0012$, $e=0.02$ and  three values of $v'(r_h)$. 
As is obvious from Fig. \ref{fig:susy1} (right) decreasing $v'(r_h)$ leads to a decrease in $\Omega$ and hence an increase in the effective mass of the scalar field. So, not suprisingly, we see that the scalar field function oscillations
(which are more or less equal in amplitude) appear on a smaller interval of the radial coordinate $r$, i.e. are stronger localized when moving along the branch C. At the same time, the minimal value of $N(r)$ decreases and gets closer to zero and accordingly, the amplitude of the Ricci scalar $R$ increases. Hence, the solution has a stronger spatial variation in the scalar curvature when moving along the branch.

\section{Conclusions}
In this paper, we have studied a novel type of scalar hair which appears for spherically symmetric black hole solutions coupled minimally to a U(1) gauge field and a self-interacting complex valued scalar field. This scalar hair appears for sufficiently large gravitational coupling and sufficiently small gauge coupling. The subtle non-linear interaction between the curvature of space-time, the scalar field and the electric field than leads to spatial oscillations of the scalar field.
Considering our results, it appears that the solution is a result of the existence of a black hole horizon and a 
``barrier'' at spatial infinity which comes from the effective mass of the scalar field. The decrease of the
horizon electric field pushes the electric charge into the Q-cloud outside the horizon such that at the end of the branch the horizon electric charge of the black hole is essentially zero. The electrically charged and gravitating cloud 
then forms a shell outside the horizon that gets ``squeezed'' between the horizon and the barrier due to the scalar field effective mass and consequently develops oscillations. These oscillations are also present in the curvature scalars such as the Ricci scalar and the Kretschmann scalar, respectively, and since they appear well outside the horizon they might have observational consequences such as influencing the motion of objects in the vicinity of the black hole. This is currently under investigation and will be discussed elsewhere. 
As a final note, let us remark that due to the similarity of the effects of a U(1) gauge field and rotation in the context of the synchronization condition, it would be interesting to see whether the scalar field oscillations are also present for the solutions studied in \cite{Herdeiro_2014}. If so, there might be a new observational Ansatz to observe black holes
with scalar hair.



 \end{document}